# Globular Clusters in the Magellanic Clouds.
# I: BV CCD-Photometry for 11 Clusters *

C.E. Corsi[1], R. Buonanno[1]
F. Fusi Pecci[2], F.R. Ferraro[2]
V. Testa[3], L. Greggio[3]

[1] Osservatorio Astronomico di Monte Mario, Roma, Italy
[2] Osservatorio Astronomico, Bologna, Italy
[3] Dipartimento di Astronomia, Bologna, Italy

Send proofs to:

F. Fusi Pecci
Osservatorio Astronomico
Via Zamboni 33
I - 40126 BOLOGNA
ITALY
37907::flavio
flavio@alma02.astro.it
fax: 39-51-259407

* Based on observations obtained at the European Southern Observatory, La Silla, Chile.

**Summary**

We present here BV CCD-data for 11 intermediate-age LMC clusters, and precisely: NGC 1756, 1831, 1868, 1987, 2107, 2108, 2162, 2173, 2190, 2209, 2249.

Though statistical sampling, field contamination, and crowding problems have made the analysis and discussion very hard to accomplish, the observational data essentially confirm the existence of the predicted *RGB phase-transition* (Renzini and Buzzoni 1986).

In particular, from the CMDs of the 11 LMC clusters down to V$\sim$22 we can conclude that:

1. In the ($V_{TO}$ , $V_{Cl,m}$ ) plane, the models yield a very good overall description of the data. A similar agreement can also be found in the ($V_{TO}$ , $\Delta(V_{TO} - V_{Cl,m})$) plane, where the ordinate is moreover distance and reddening independent.
2. With the current sample, it is still impossible to firmly choose between "classical" and "overshooting" models. Both sets yield a good fit to the data in luminosity, classical models being apparently better if the observational results are taken at face value.
3. Regardless of the adopted distance modulus and reddening, the separation in colour between the MS-band (H-burning) and the Red Clump (He-burning) is smaller than predicted by any theoretical tracks, either classical or with overshooting. In particular, the MS is too red by about 0.05–0.10 mag and the Red Clump is more extended than expected.
4. The existence of the so-called RGB phase transition seems to be confirmed. In particular, the behaviour of the luminosity of Red Clump stars and the RGB development are qualitatively consistent with the theoretical predictions. Finally, we have identified a small sub-set of clusters (NGC 2209, 2190, 2162) to pick up for future deeper study (maybe with HST) and which are most suitable for a further detailed investigation on this subject.

## 1. INTRODUCTION

It is now widely recognized that the globular clusters (GCs) of the Magellanic Clouds (MC) offer a unique tool for testing several predictions of stellar evolution theory, as well as for sharpening our understanding of the evolution of the integrated properties of stellar populations (van den Bergh 1981, hereafter vdB81; Renzini 1981, 1991; Renzini and Buzzoni 1986 –RB86; Bica, Dottori and Pastoriza 1986, –BDP86; Chiosi, Bertelli, and Bressan 1988 –CBB88; Brocato *et al.* 1989; Battinelli and Capuzzo Dolcetta 1989; Alongi and Chiosi 1989; Frogel, Mould and Blanco 1990 –FMB90; Meurer, Cacciari, and Freeman 1990 –MCF90; Barbero *et al.* 1990; Barbaro and Olivi 1991; Mould, 1992; Arimoto and Bica, 1989; Bica *et al.* 1991 –BCDSP91; Bica, Claria, and Dottori 1992; Bressan, Chiosi and Fagotto 1993 –BCF93; Girardi and Bica 1993, and the proceedings edited by Chiosi and Renzini 1986, Kron and Renzini 1988, Haynes and Milne 1991, Barbuy and Renzini 1992, Smith and Brodie 1993, and references therein).

These goals require an appropriate ranking of clusters with varying ages and metallicities and the knowledge of the detailed morphology of their Colour-Magnitude Diagrams (CMDs) and Luminosity Functions (LFs) based on accurate photometry, possibly carried out from the ultraviolet to near IR bands.

Among the various topics open to investigations, we had initially singled out one specific aspect: the origin of the bimodal distribution of the integrated B−V colours of MC clusters (Gascoigne and Kron, 1952; Gascoigne, 1971,1980; Searle, Wilkinson and Bagnuolo 1980 –SWB; vdB81; Renzini 1981, 1992; RB86; Elson and Fall 1985,1988 [–EF85 and EF88]; CBB88; BCDSP91; BCF93).

The motivations for this choice are manifold but they are part of a unique strategy aimed at using the MC clusters as *template* stellar populations for studying high redshift (elliptical) galaxies for cosmological purposes (Renzini 1991; Chambers and Charlot 1990; BCDSP91; Bruzual and Charlot 1993; BCF93).

Very schematically (see for instance the discussion in RB86), if the cluster integrated colour variations could be strictly correlated to known evolutionary time-scales of well identified cluster members, and if at least some of the primeval galaxies could be considered to be formed by "simple" stellar populations (i.e. coeval and with small metallicity spread, as in clusters), then the most evident observed colour glitches could be used as "calibrated clocks" in the study of the epoch of galaxy formation.

Concerning specifically the integrated colour bimodality of the MC GCs, as repeatedly noticed and shown for instance in Fig. 13 of RB86, a plot of the integrated B−V versus the types defined by SWB reveals that the integrated colour transition takes place within the SWB-type IV. Therefore, we have mainly concentrated our observational efforts on clusters of this class, to determine the current evolutionary phases of the stars from which the observed integrated colour transition originates.

The occurrence of such a transition within SWB-class IV is clearly indicated in particular by the two-colour diagrams –(U-B,B-V)– presented by EF85 (Fig. 1) and recently by BCDSP91 (Fig. 1). An even more stringent evidence for this comes from the (U-B,V-K) two-colour diagram here shown in Fig. 1 (see also Renzini 1991, Fig. 3). In fact, from an inspection of the plot, it can be seen that while the clusters belonging to the other SWB-classes occupy sufficiently well defined areas, those of SWB-class IV are spread out over the total range of observed V-K colours. Since the corresponding spread in the U-B colours for these clusters is quite small, this evidence indicates that the colour transition is much more evident when considering redder bands. Hence, one may conclude that red (cool) stars are probably responsible for its origin.

Several possible causes for the quoted colour transition have been proposed so far, *i.e.* (i) the so-called "Asymptotic Giant Branch and/or Red Giant Branch phase transitions" (Gascoigne, 1971,1980; Renzini, 1981; Renzini and Buzzoni, 1983; RB86); (ii) an age gap and/or effects of cluster disruption (vdB81); (iii) a peculiar age-metallicity relation inducing a hook in the distribution in the two-colour diagrams (Frenk and Fall, 1982), and, (iv) finally, the result of the combination of different effects (CBB88, BCF93 –age being probably the most important–, Battinelli and Capuzzo Dolcetta 1989).

With the studies carried out and presented in this series of papers we aim in particular at checking observationally the idea originally proposed by Renzini (1981) and RB86 that this steep integrated colour change occuring in the MC clusters of SWB-type IV may be originated by the so-called "Red Giant Branch phase-transition" (hereafter *RGB ph-t* ).

More specifically, the essence of the claim by RB86 is based on the model prediction first stressed by Iben (1967) that a major dichotomy exists in the properties of the RGB evolution between stars of low ($M<2.25M_{\odot}$) and intermediate ($2.25M_{\odot} <M< 8M_{\odot}$) mass. The development of the RGB (the portion of the Hydrogen-shell-burning phase spent close to the Hayashi track) occurs only if the star is less massive than a critical value (hereafter $M_{HeF}$ ) which separates core Helium ignition in degenerate ($M_i<M_{HeF}$ ) or non-degenerate ($M_i>M_{HeF}$ ) conditions. The evolution of stars of initial mass around $M_{HeF}$ has been purposely investigated by Sweigart, Greggio and Renzini (1989,1990) through the computation of a fine grid of sequences with standard input physics. These models show that the development of an extended RGB should occur rather abruptly at an age of approximately 0.6 Gyr, almost independently of chemical composition. Hence, as soon as stars of the appropriate, critical initial mass start evolving off the Main Sequence, the *sudden* appearance of bright and red RGB stars would induce a steep integrated colour variation of the global population. In the RB86 framework (see their Fig. 5), a similar colour glitch could be originated by the first appearance in the population of Asymptotic Giant Branch stars (see also Gascoigne 1971,1980), and this feature too should be somehow detectable in the red and IR colours.

This overall picture has been recently revised by Renzini (1992), following the results of new evolutionary computations made by Blocker and Schönberner (1991). They show that, while experiencing the envelope burning, the more massive AGB stars climb quickly up to very high luminosities, where severe mass loss is likely to interrupt their evolution along the AGB. Correspondingly, the AGB phase transition is delayed until the mass of the evolving star is too low to experience the envelope burning process. As a result, the ages at which the AGB and RGB phase transitions occur become closer, and the V-K colour jump can be ascribed to a combination of the AGB+RGB development. Besides, since these stars radiate mostly in the IR, the effect on the integrated B-V colours is expected to be modest.

In this respect, the RB86 working hypothesis has been tested and questioned via models and simulations, for instance by CBB88, BCF93. In particular, BCF93 conclude: "the phase transition (either AGB or RGB) cannot explain the gap of about 0.3 mag observed in the distribution of the (B-V) colour of LMC clusters or equivalently in the relation between the cluster SWB-type and (B-V). Instead, following CBB88, we attribute the gap to the complicated history of cluster formation and disruption that took place in the LMC".

From the above discussion, it is quite evident that the best direct test is the quantitative analysis of the observed CMDs. Before presenting the results however, we have to stress immediately three crucial items:

i) Since the LMC clusters have absolute integrated luminosities of a few $10^4 L\odot$ and the AGB and bright-RGB lifetimes are quite short ($\leq 10^7 yr$), statistical fluctuations will dominate the counts due to the intrinsic poorness of the samples of AGB and RGB stars expected in a single cluster. This implies that many clusters should be observed and the samples properly added.

ii) Most of the LMC clusters are projected on a crowded background, and field contribution due to LMC stars having similar or different ages can strongly affect the counts. A proper description of the CMD properties of the LMC underlying population is therefore necessary.

iii) The B,V photometric bands may not be the most appropriate to evidence the actual contribution of stars as red as the AGB and RGB objects. For this reason we have also undertaken a parallel study in the JHK infrared bands, whose results are presented in a companion paper (Ferraro *et al.* 1993, hereafter Paper II).

Here we present a first set of data obtained from the BV CCD-photometry of a sample of 11 clusters in the Large Magellanic Cloud (LMC). More precisely, we deal with NGC 1756 (SWB-type III), 1831 (V), 1868 (IV), 1987 (IV), 2107 (IV), 2108 (IV-V), 2162 (V), 2173 (V-VI), 2190 (IV-V), 2209 (III-IV), 2249 (IV). They were selected at the beginning of the project by choosing a sub-set of the objects reported in the (U-B,B-V) diagram of EF85 and located in the region corresponding substantially to SWB-class IV, with $s = 31-45$.

For each cluster we report the results of the photometric survey we carried out using the ESO telescopes. The basic aim of the observations was to get a preliminary general morphology of the main branches in the CMDs for a very wide sample of clusters in order to pick up a smaller sub-set including the most suitable clusters for the investigation of the quoted *RGB ph-t*. As stated, Paper II of this series (Ferraro *et al.* 1993) is devoted to present the results of a similar survey carried out for the same clusters in JHK at CTIO with an IR-array. Moreover, since BCDSP91 have meanwhile presented a new list of MC clusters specifically crucial for studying the AGB phase transition, we plan to insert a subsample of the clusters listed in their Table 1 A-B in our observing material to make our analysis sharper and more complete. Future papers will then report on the next steps of the observations, a *quantitative* treatment of the CMDs and the LFs for a subset of important clusters, and a complete discussion.

**Table 1.** Log of the observations.

| Cluster | Run | Telescope | Filter | $N_B$ | $N_V$ | Min. | Max. exp. (s) | $< FWHM >$ |
|---|---|---|---|---|---|---|---|---|
| NGC 1756 | 2,4 | 2.2 MPI | 279,280 | 4 | 4 | 15 | 1320 | 1.1" |
| | | | 445,446 | | | | | |
| NGC 1831 | 2,4,5 | 2.2 MPI | 279,280 | 4 | 4 | 10 | 1500 | 1.3" |
| | | 3.5 NTT | 445,446 | 1 | 1 | 300 | 900 | 2.4" |
| NGC 1868 | 2,4 | 2.2 MPI | 279,280 | 3 | 3 | 120 | 1500 | 1.5" |
| | | | 445,446 | | | | | |
| NGC 1987 | 2,4,5 | 2.2 MPI | 279,280 | 3 | 3 | 15 | 1200 | 1.6" |
| | | 3.5 NTT | 445,446 | 1 | 1 | 300 | 900 | 3.0" |
| NGC 2107 | 2,4 | 2.2 MPI | 279,280 | 3 | 3 | 60 | 1500 | 1.8" |
| | | | 445,446 | | | | | |
| NGC 2108 | 4 | 2.2 MPI | 445,446 | 2 | 2 | 60 | 480 | 1.5" |
| NGC 2162 | 3,4,5 | 2.2 MPI | 445,446 | 2 | 2 | 900 | 2100 | 1.1" |
| | | 3.5 NTT | | 1 | 1 | 300 | 900 | 2.8" |
| NGC 2173 | 1,4,5 | 1.5 Danish | 445,446 | 1 | 1 | 1800 | 3600 | 1.6" |
| | | 2.2 MPI | | | | | | |
| | | 3.5 NTT | | 1 | 1 | 180 | 300 | 1.8" |
| NGC 2190 | 3,4 | 2.2 MPI | 445,446 | 3 | 3 | 180 | 1500 | 1.2" |
| NGC 2209 | 2,4 | 2.2 MPI | 279,280 | 4 | 4 | 60 | 1500 | 0.9" |
| | | | 445,446 | | | | | |
| NGC 2249 | 2,4 | 2.2 MPI | 279,280 | 1 | 1 | 600 | 1200 | 1.3" |
| | | | 445,446 | | | | | |

Runs: **1** – *26-28/10/1984*, **2** – *7-12/12/1985*, **3** – *2-5/12/1986*, **4** – *12-17/12/1987* , **5** – *13/11/1990*

chip: ESO # 5 RCA (runs 1, 2, 3, 4), Tektronix 1024×1024 (run 5)

**Table 2.** Internal photometric errors.

| Cluster | No. frames | No. stars measured | $V < 19.5$ $\sigma(V)$ | $V < 19.5$ $\sigma(B-V)$ | $V > 19.5$ $\sigma(V)$ | $V > 19.5$ $\sigma(B-V)$ |
|---|---|---|---|---|---|---|
| NGC 1756 | 8 | 803 | 0.01 | 0.02 | 0.01 | 0.03 |
| NGC 1831 | 8 | 1417 | 0.02 | 0.03 | 0.03 | 0.04 |
| NGC 1868 | 6 | 1448 | 0.01 | 0.03 | 0.03 | 0.05 |
| NGC 1987 | 6 | 1655 | 0.03 | 0.05 | 0.05 | 0.08 |
| NGC 2107 | 6 | 1303 | 0.01 | 0.01 | 0.02 | 0.03 |
| NGC 2108 | 4 | 789 | 0.01 | 0.04 | 0.02 | 0.06 |
| NGC 2162 | 6 | 851 | 0.01 | 0.01 | 0.01 | 0.02 |
| NGC 2173 | 2 | 616 | 0.01 | 0.05 | 0.03 | 0.08 |
| NGC 2190 | 6 | 971 | 0.01 | 0.01 | 0.01 | 0.02 |
| NGC 2209 | 6 | 1177 | 0.01 | 0.03 | 0.02 | 0.04 |
| NGC 2249 | 2 | 391 | 0.01 | 0.03 | 0.02 | 0.04 |

**Table 3.** Photometric data from literature.

| Cluster | Other names | $V_{int}$ | $(B-V)_{int}$ | $(U-B)_{int}$ | $E(B-V)$ | $SWB$ | $s$ |
|---|---|---|---|---|---|---|---|
| NGC 1756 | SL 94 | 12.24[2] | 0.40[2] | 0.09[2] | | | 32[2] |
| NGC 1783 | SL 148 | 10.93[2] | 0.62[2] | 0.23[2] | 0.10[11] | V[1,6] | 37[4] |
| | | | | | 0.06[35] | | 38.0[11] |
| NGC 1806 | SL 184 | 11.10[2] | 0.73[2] | 0.26[2] | 0.12[3] | V[1] | 40[4] |
| NGC 1831 | SL 227 | 11.18[2] | 0.34[2] | 0.13[2] | 0.10[3] | V[1] | 31[4] |
| | LW 133 | 10.59[28] | 0.35[28] | | 0.05[39] | | 32.7[11] |
| | | | | | 0.04[35] | | |
| | | | | | 0.07[11] | | |
| NGC 1868 | SL 330 | 11.56[2] | 0.45[2] | 0.15[2] | 0.07[3] | IV[6,7] | 33[4] |
| | LW 169 | | | | | | 34.5[11] |
| | ESO 085-SC56 | | | | | | |
| NGC 1978 | SL 501 | 10.70[2] | 0.78[2] | 0.23[2] | 0.10[3] | VI[1] | 45[4] |
| | | | | | 0.07[11] | | 43.8[11] |
| | | | | | 0.19[11] | | |
| NGC 1987 | SL 486 | 12.08[2] | 0.52[2] | 0.20[2] | 0.12[3,14] | IV[1] | 35[4] |
| | | 11.50[14] | | | | | 35.1[11] |
| NGC 2107 | SL 679 | 11.51[2] | 0.38[2] | 0.13[2] | 0.19[3] | IV[1] | 32[4] |
| NGC 2108 | SL 686 | 12.32[2] | 0.58[2] | 0.22[2] | 0.18[3] | IV – V[7] | 36[4] |
| NGC 2162 | SL 814 | 12.70[2] | 0.68[2] | 0.31[2] | 0.07[3] | V[1] | 39[4] |
| | | | | | 0.05[11] | | 40.5[11] |
| | | | | | 0.04[23] | | |
| | | | | | 0.06[24] | | |
| NGC 2173 | SL 807 | 12.30(62")[2] | 0.84[2] | 0.34[2] | 0.07[3,11] | V – VI[1] | 42[4] |
| | LW 348 | 13.28(25")[2] | 0.86[2] | 0.50[2] | 0.12[25] | VI[6,7] | 42.5[11] |
| NGC 2190 | SL 819 | | | | 0.10[23] | | |
| | LW 357 | | | | | | |
| | ESO033-SC36 | | | | | | |
| NGC 2209 | SL 849 | 13.15[2] | 0.53[2] | 0.20[2] | 0.07[3] | III – IV[1] | 35[4] |
| | LW 408 | | | | 0.15[38] | | 36.9[11] |
| | | | | | 0.06[11] | | |
| NGC 2249 | SL 893 | 12.23[2] | 0.43[2] | 0.20[2] | 0.12[11] | | 34[4] |
| | LW 479 | 12.17(100")[8] | 0.39[8] | 0.21[8] | 0.10[32] | | 33.6[11] |
| | | 11.94(150")[8] | 0.42[8] | 0.20[8] | | | |

References: see Table 4

**Table 4.** Astrophysical data from literature.

| Cluster | $t_8(op)$ | $t_8(ir)$ | [Fe/H] | $M_{tot} \times 10^5 (M\odot)$ | $r_c$ | $r_t$ | $v_r$ |
|---|---|---|---|---|---|---|---|
| NGC 1756 | 3.5[4] | | | | | | |
| | 3.8[5] | | | | | | |
| NGC 1783 | 33±3[7] | < 30[14] | -0.9±0.4[7] | 3[30] | 4.9±0.4 pc | | 274[40] |
| | 9[27] | | -0.45±0.3[18] | | | | 277[6] |
| | 16[30] | | | | | | |
| | 11[12] | | | | | | |
| | 25[10a] | | | | | | |
| | 7.9[10b] | | | | | | |
| NGC 1806 | 43±3[7] | < 30[14] | -0.23[40] | 0.9[33] | 3.7pc[33] | 58.8pc[33] | 225[40] |
| | | < 40[15] | -0.7±0.35[7] | | | | 220±10[6] |
| NGC 1831 | 25.7[11] | < 25[14] | 0.01[40] | 0.4[34] | 11.8"[35] | 187"[35] | 280[40] |
| | 3.5cl[39] | < 40[15] | -0.33[39] | | 5.4pc[34] | 54pc[34] | 253±13[6] |
| | 5.5ov[39] | | -0.1[29] | | | | |
| | 4[21] | | -1.2[18] | | | | |
| | 6.3[10a] | | | | | | |
| | 25[10b] | | | | | | |
| | 5[29] | | | | | | |
| NGC 1868 | 5[8,11,7] | 7[16] | -0.50[40] | | 6.1"[37] | | 283[40] |
| | 7[26] | | -0.6±0.35[7] | | | | 260±30[6] |
| | 3.3[5] | | -1.2[9] | | | | |
| | 10[9a] | | | | | | |
| | 17.8[9b] | | | | | | |
| | 13.5[9c] | | | | | | |
| NGC 1978 | 21[12,26] | < 60[13] | -0.5±0.2[18] | 3[33] | 3.0pc[28] | | 293.3[40] |
| | 25.1[11] | < 15[14] | -0.7[12] | | | | 286±8[6] |
| | 20[26,30] | < 20[15] | -0.42[40] | | | | 293±3[41] |
| | 66[7] | | -1.1[7] | | | | |
| | 19.9[9a] | | | | | | |
| | 14.1[9b] | | | | | | |
| | 17.8[9c] | | | | | | |
| | 12.2–19.9[9d] | | | | | | |
| NGC 1987 | 8±3[7] | < 25[14] | -1.0±0.3[7] | | 2.9±0.3pc[47] | | 253±23[6] |
| | 15[4] | < 30[15] | | | | | |
| | 4.7[11] | | | | | | |
| NGC 2107 | 4[37] | < 10[14] | | 0.9[34] | 3.4±0.4pc[37] | 54pc[34] | 248±13[6] |
| | | < 15[15] | | | 5.4pc[34] | | |
| NGC 2108 | 7.9[37] | | -1.2±0.2[7] | | 2.5±0.4pc[37] | | |
| | 22±3[7] | | | | | | |
| NGC 2162 | 38±4[7] | < 10[14] | -0.23[24,40] | | | | 322[40] |
| | 7.41[11] | < 11[15] | -0.2[23] | | | | |
| | 10[23,24] | | -1.35±0.3[7] | | | | |
| | 15.8[10a] | | -1.2[10] | | | | |
| | 12.6[10b] | | | | | | |

**Table 4.** continue.

| Cluster | $t_8(op)$ | $t_8(ir)$ | [Fe/H] | $M_{tot} \times 10^5 (M\odot)$ | $r_c$ | $r_t$ | $v_r$ |
|---|---|---|---|---|---|---|---|
| NGC 2173 | 21±4[7] | > 50[13] | -1.4±0.2[7] | 0.5[34] | 6.2–1.9pc[34] | 62pc[34] | 241[40] |
| | 65±7[7] | < 100[15] | -0.24[6] | | | | 232±22[6] |
| | 15.1[11] | | -0.75±0.4[25] | | | | |
| NGC 2190 | 10[23] | < 40[13] | -0.12[40] | | | | 260[40] |
| | 12.6[9a] | < 25[14] | -1.2[9] | | | | |
| | 39.8[9b] | < 30[15] | -0.2[23] | | | | |
| | 22.4[9c] | | | | | | |
| | 17.4–26.3[9d] | | | | | | |
| NGC 2209 | 8.4±2[8] | < 40[13] | -1.2[17] | | 5.0pc[36] | | 255[40] |
| | 11[11] | < 20[14] | -0.9±0.3[7] | | | | |
| | 10[36] | < 30[15] | -1[19] | | | | |
| | 7±1[5] | | | | | | |
| | 12±2[38] | | | | | | |
| | 12±3[7] | | | | | | |
| | 8[22] | | | | | | |
| | 15.9[9a] | | | | | | |
| | 20.9[9b] | | | | | | |
| | 17.8[9c] | | | | | | |
| | 15.8–25.1[9d] | | | | | | |
| NGC 2249 | 5.5±1.5[8,31] | | $.002 < Z < .015$[31] | | | | |
| | 6[11,31] | | | | | | |
| | 7[31] | | | | | | |

Notes: column 6: 1pc at $(m - M)_0 = 18.5$ corresponds to 4.12 arcsec.

References: (1)- SWB, 1980; (2)- van den Bergh, 1981; (3)- Persson *et al.*, 1983; (4)- Elson and Fall, 1985; (5)- Elson and Fall, 1988; (6)- Freeman, Illingworth and Oemler, 1983; (7)- Bica, Dottori and Pastoriza, 1986; (8)- Bica *et al.*, 1991; (9)- Chiosi *et al.*, 1986; (10)- Chiosi, Bertelli and Bressan, 1988; (11)- Meurer, Cacciari and Freeman, 1990; (12)- Frogel, Mould and Blanco, 1990; (13)- Mould and Aaronson, 1980 (AMMA I); (14)- Aaronson and Mould, 1982 (AMMA II); (15)- Mould and Aaronson, 1982 (AMMA III); (16)- Aaronson and Mould, 1985 (AMMA IV); (17)- Gascoigne, 1980; (18)- Cohen, 1982; (19)- Rabin, 1982; (20)- Hodge, 1983; (21)- Hodge, 1984; (22)- Flower, 1984; (23)- Schommer, Olszewski and Aaronson, 1986; (24)- Chiosi and Pigatto, 1986; (25)- Mould, Da Costa and Wieland, 1986; (26)- Mould and Da Costa (1988); (27)- Mould *et al.*, 1989; (28)- Mateo, 1987; (29)- Mateo, 1988; (30)- Mateo, 1992; (31)- Jones, 1987; (32)- Burstein and Heiles, 1982; (33)- Kontizas, Chrysovergis and Kontizas, 1987; (34)- Chrysovergis, Kontizas and Kontizas, 1989; (35)- Westerlund, 1990; (36)- Elson, 1991; (37)- Elson, 1992; (38)- Dottori *et al.*, 1987; (39)- Vallenari *et al.*, 1992; (40)- Olszewski *et al.*, 1991; (41)- Fischer, Welch and Mateo, 1992.

Ref. 10: (a)- AGB star tip luminosity; (b)- MS-fitting with overshoot.
Ref. 9: (a)- MS-fitting with overshoot; (b)- red clump luminosity; (c)- coincidence red clump – MS; (d)- AGB tip luminosity, with various mass loss parametrizations.

**Table 5.** MW foreground.

| | 13–15 | 15–17 | 17–19 | 19–21 | 21–23 |
|---|---|---|---|---|---|
| $B-V<0.8$ | 0.041–0–3 | 0.086–1–5 | 0.091–1–6 | 0.190–1–12 | 0.180–1–11 |
| $0.8<B-V<1.3$ | 0.16–0–1 | 0.078–0/1–5 | 0.160–1–10 | 0.130–1–8 | 0.220–1–13 |
| $1.3<B-V$ | 0.003–0–2 | 0.017–0–1 | 0.110–1–7 | 0.390–2–24 | 0.870–5–53 |

Notes: for each box, first number is No. of stars per square arcmin, second number is No. of stars expected over an area covered by CCD RCA + 2.2 Telescope, third number is No. of stars expected over an area covered by EMMI+NTT.
Area covered by: RCA+2.2 = 6 square arcmin, RCA+1.5 = 10 square arcmin, 3.5NTT = 61.4 square arcmin.

Reference: Ratnatunga and Bahcall, 1985.

**Table 17a.** Mean Loci from CMDs.

| Cluster | $V_{TO}$ | $(B-V)_{TO}$ | $< V_{Cl} >$ | $< (B-V)_{Cl} >$ | $V_{Cl,m}$ | $(B-V)_{Cl,m}$ | $\Delta V_{Cl}$ | $\Delta(B-V)_{Cl}$ |
|---|---|---|---|---|---|---|---|---|
| NGC 1756 | 17.6 | 0.02 | 17.3 | 0.90 | 18.0 | 1.25 | 16.5–18.0 | 0.45–1.40 |
| NGC 1831 | 18.3 | 0.10 | 18.5 | 0.90 | 19.0 | 0.95 | 18.2–19.0 | 0.65–1.00 |
| NGC 1868 | 19.3 | 0.15 | 19.3 | 0.75 | 19.7 | 0.85 | 18.8–19.7 | 0.65–0.85 |
| NGC 1987 | 18.8 | 0.18 | 19.2 | 0.85 | 19.5 | 0.90 | 18.7–19.5 | 0.75–0.95 |
| NGC 2107 | 18.0 | 0.15 | 17.5 | 1.00 | 18.8 | 1.05 | 17.0–18.8 | 0.50–1.20 |
| NGC 2108 | 19.1 | 0.20 | 19.3 | 0.90 | 19.8 | 0.87 | 18.8–19.8 | 0.80–1.10 |
| NGC 2162 | 19.6 | 0.25 | 19.2 | 0.87 | 19.4 | 0.90 | 18.8–19.4 | 0.80–1.00 |
| NGC 2173 | 20.0 | 0.40 | 19.1 | 0.87 | 19.3 | 0.87 | 18.7–19.3 | 0.80–0.95 |
| NGC 2190 | 19.5 | 0.20 | 19.5 | 0.86 | 19.8 | 0.92 | 18.7–19.8 | 0.80–0.95 |
| NGC 2209 | 19.5 | 0.25 | 19.7 | 0.88 | 20.0 | 0.95 | 19.2–20.0 | 0.80–1.00 |
| NGC 2249 | 18.9 | 0.18 | 18.8 | 0.90 | 19.3 | 0.87 | 18.3–19.3 | 0.80–1.00 |

**Table 17b.** Mean Loci of field CMDs.

| Cluster | $V_{TO}$ | $(B-V)_{TO}$ | $< V_{Cl} >$ | $< (B-V)_{Cl} >$ | $\Delta V_{Cl}$ | $\Delta(B-V)_{Cl}$ | Notes |
|---|---|---|---|---|---|---|---|
| NGC 1756 | 16.4 | 0.05 | | | | | young |
| | 19.5 | 0.15 | 19.3 | 0.95 | 18.9–19.7 | 0.80–1.10 | inter.–old |
| NGC 1831 | 18.8 | 0.15 | 18.8 | 0.85 | 18.2–19.3 | 0.80–1.00 | sim.to cluster |
| | 19.0 | 0.10 | 19.0 | 0.9 | 18.7–19.3 | 0.80–1.00 | NTT field |
| NGC 1868 | 19.7 | 0.17 | 19.2 | 0.80 | 18.7–19.6 | 0.75–0.90 | sim.to cluster |
| NGC 1987 | 19.5 | 0.20 | 19.3 | 0.90 | 18.9–19.6 | 0.80–1.05 | |
| | 15.0 | -0.20 | | | | | NTT field-very young |
| NGC 2107 | 19.8 | 0.25 | 19.4 | 0.95 | 19.0–19.8 | 0.80–1.10 | |
| NGC 2108 | 16.7 | -0.03 | | | | | young |
| | 19.5 | 0.15 | 19.5 | 0.97 | 19.0–20.0 | 0.85–1.15 | interm.-old |
| NGC 2162 | 20.0 | 0.25 | 19.3 | 0.90 | 19.0–19.5 | 0.80–1.00 | |
| | 20.6 | 0.35 | 19.2 | 0.90 | 18.8–19.4 | 0.75–1.05 | NTT field |
| NGC 2173 | 20.4 | 0.40 | 19.1 | 0.90 | 19.0–19.2 | 0.8–1.0 | HB |
| | 19.0 | 0.05 | 19.1 | 0.90 | 18.7–19.7 | 0.75–1.15 | NTT field-interm. |
| NGC 2190 | 19.7 | 0.15 | 19.3 | 0.85 | 18.9–19.5 | 0.80–1.00 | |
| NGC 2209 | 19.8 | 0.30 | 19.4 | 0.92 | 19.2–19.9 | 0.85–1.00 | |
| NGC 2249 | 19.7 | 0.18 | 18.6 | 0.90 | 18.2–19.0 | 0.75–1.05 | poorly populated |

**Table 18.** Mean Loci from literature.

| Cluster | Reference | $V_{TO}$ | $(B-V)_{TO}$ | $< V_{Cl} >$ | $< (B-V)_{Cl} >$ | $\Delta V_{Cl}$ | $\Delta(B-V)_{Cl}$ | Notes |
|---|---|---|---|---|---|---|---|---|
| NGC 1831 | Hodge 84 | 19.3 | 0.00 | 18.4 | 0.80 | 18.0–18.8 | 0.70–0.90 | field not subtracted–blending |
| NGC 1831 | ” | 19.3 | 0.00 | 18.2 | 0.80 | 18.0–18.5 | 0.70–0.90 | field subtracted |
| NGC 1831 | ” | 18.9 | 0.00 | 19.1 | 0.80 | 18.9–19.3 | 0.65–1.00 | field CMD |
| NGC 1831 | Vallenari *et al.* 92 | 18.3 | 0.05 | 18.5 | 0.75 | 18.0–19.0 | 0.65–0.9 | |
| NGC 1868 | Flower *et al.* 80 | 19.3 | 0.10 | 19.5 | 0.70 | 18.7–20.0 | 0.60–0.80 | |
| NGC 1868 | ” | ∼18.8 | ∼0.10 | ∼19.5 | ∼0.75 | 19.0–20.0 | 0.70–0.80 | field CMD–too sperse and poor |
| NGC 2162 | Schommer *et al.* 84 | 19.4 | 0.25 | 19.1 | 0.76 | 18.7–19.6 | 0.70–0.88 | |
| NGC 2190 | Schommer *et al.* 84 | 19.4 | 0.23 | 19.5 | 0.88 | 18.9–19.8 | 0.75–0.96 | |
| NGC 2209 | Dottori *et al.* 87 | 19.5 | 0.24 | 19.3 | 0.75 | 19.0–20.0 | 0.65–0.95 | |
| NGC 2209 | Gascoigne 76 | 19.4 | 0.28 | 19.2 | 0.80 | 19.0–19.8 | 0.70–0.95 | photographic |
| NGC 2249 | Jones 87 | 19.2 | 0.15 | 19.2 | 0.85 | | | very few dense-field not subtracted |
| NGC 2249 | ” | 19.3 | 0.15 | 19.0 | 0.82 | 18.6–19.5 | 0.10–1.00 | field subtracted |
| NGC 2249 | ” | 19.4 | 0.17 | 19.2 | 0.83 | 18.7–19.4 | 0.78–0.96 | field CMD |